\newlength{\absize}
\documentclass[12pt]{article}
\usepackage{latexsym}
\usepackage{amssymb}
\usepackage{setspace}
\newcommand{\figsize}{\small}

\input prepictex
\input pictex
\input postpictex
\newdimen\tdim
\tdim=\unitlength
\def\stpltsmbl{\setplotsymbol ({\small .})}

\def\tarrow{\arrow <5\tdim> [.3,.6]}

\newbox\sru
\setbox\sru=\hbox{\beginpicture
\setcoordinatesystem units <\tdim,\tdim>
\stpltsmbl
\setquadratic
\plot
  0.0   0.0
  4.8   1.5
  7.5   5.0
  7.3   8.5
  5.0  10.0
  2.7   8.5
  2.5   5.0
  5.2   1.5
 10.0   0.0
/
\endpicture}
\def\springru #1 #2 *#3 /{\multiput {\copy\sru}  at
#1 #2 *#3 10 0 /}

\renewcommand{\bar}{\overline}

\newcommand{\spur}[1]{\!\not\! #1 \,}
\newcommand{\un}{\mathcal{U}}

\newcommand{\cO}{\mathcal{O}}
\renewcommand{\slash}[1]{#1\!\!\!/}

\newcommand{\be}{\begin{equation}}
\newcommand{\ee}{\end{equation}}
\newcommand{\bea}{\begin{eqnarray}}
\newcommand{\eea}{\end{eqnarray}}
\newcommand{\nn}{\nonumber}

\usepackage{epsf}

\def\ltilde#1{\mathord{\mathop{\kern 0pt #1}\limits_{\sim\atop}}}

\begin{document}

\thispagestyle{empty}
\pagestyle{empty}
\newcommand{\starttext}{\newpage\normalsize
 \pagestyle{plain}
 \setlength{\baselineskip}{3ex}\par
 \setcounter{footnote}{0}
 \renewcommand{\thefootnote}{\arabic{footnote}}
 }
\newcommand{\preprint}[1]{\begin{flushright}
 \setlength{\baselineskip}{3ex}#1\end{flushright}}
\renewcommand{\title}[1]{\begin{center}\LARGE
 #1\end{center}\par}
\renewcommand{\author}[1]{\vspace{2ex}{\Large\begin{center}
 \setlength{\baselineskip}{3ex}#1\par\end{center}}}
\renewcommand{\thanks}[1]{\footnote{#1}}
\renewcommand{\abstract}[1]{\vspace{2ex}\normalsize\begin{center}
 \centerline{\bf Abstract}\par\vspace{2ex}\parbox{\absize}{#1
 \setlength{\baselineskip}{2.5ex}\par}
 \end{center}}

\preprint{}
\title{An Unparticle Example in 2D}
\author{
 Howard~Georgi,\thanks{\noindent \tt georgi@physics.harvard.edu}
 Yevgeny~Kats\,\thanks{\noindent \tt kats@physics.harvard.edu}
 \\ \medskip
Center for the Fundamental Laws of Nature\\
Jefferson Physical Laboratory \\
Harvard University \\
Cambridge, MA 02138
 }
\date{\today}
\abstract{We discuss what can be learned about unparticle physics by
studying simple quantum field theories in one space and one time
dimension. We argue that the exactly soluble 2D theory of a massless
fermion coupled to a massive vector boson,
the Sommerfield model, is an interesting analog of a Banks-Zaks model,
approaching a free theory at high energies and a
scale invariant theory with nontrivial anomalous dimensions at low
energies. We construct a toy standard model coupling to the fermions in
the Sommerfield model and study how the transition from unparticle
behavior at low energies to free particle behavior at high energies
manifests itself in interactions with the toy standard model particles. }

The term ``unparticle physics'' was coined by one of us to describe a
situation in which standard model physics is weakly coupled at high
energies to a sector that flows to a scale-invariant theory in
the infrared~\cite{Georgi:2007ek,Georgi:2007si}. In this class of models,
one may see surprising effects from
the production of unparticle stuff\footnote{We prefer ``unparticle stuff''
to ``unparticles'' for the physical states, because it is not clear to us
what the noun ``unparticle'' is supposed to mean, and certainly not
clear
whether it should be singular or plural.} in the scattering of standard model
particles. Studying such models forces us to confront some interesting
issues in scale invariant theories and effective field theories.

It is important to remember that unparticle physics is not just about a
scale invariant theory. There are two other important ingredients.
A crucial one is the coupling of the unparticle fields to the standard
model. Without this coupling, we would not be able to ``see'' unparticle stuff.
Also important is
the transition in the Banks-Zaks theory~\cite{Banks:1981nn} from which
unparticle physics emerges
from perturbative physics at
high energies to scale
invariant unparticle behavior at low energies. This allows us to find
well-controlled
perturbative physics that produces the coupling of the unparticle sector
to the standard model.
Without this transition, the coupling of the
standard model to the unparticle fields would have to be put in
by hand in a completely arbitrary way and much of the
phenomenological interest of the
unparticle metaphor would be lost.

In this paper, we explore the physics of the transition from unparticle
behavior at low energies to perturbative behavior at high energies in a
model with one space and one time dimension in which the analog of the
Banks-Zaks model is
exactly solvable. This will enable us to see how the transition takes place
explicitly in a simple inclusive scattering process (figure~\ref{fig-1}).
{\figsize\begin{figure}[htb]
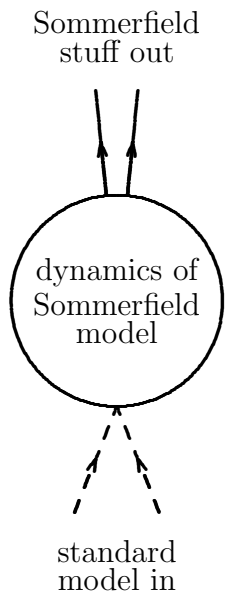

$$\beginpicture
\setcoordinatesystem units <0.8\tdim,0.8\tdim>
\stpltsmbl
\circulararc 360 degrees from 50 0 center at 0 0
\put {\stack{dynamics of,Sommerfield,model}} at 0 0
\startrotation by .995 -.1 about 0 0
\tarrow from 0 50 to 0 75
\plot 0 50 0 100 /
\stoprotation
\startrotation by .995 .1 about 0 0
\tarrow from 0 50 to 0 75
\plot 0 50 0 100 /
\stoprotation
\put {\stack{Sommerfield,stuff out}} at 0 125
\setdashes
\plot -20 -100 0 -50 /
\plot 20 -100 0 -50 /
\tarrow from -20 -100 to -10 -75
\tarrow from 20 -100 to 10 -75
\put {\stack{standard,model in}} at 0 -125
\linethickness=0pt
\putrule from -100 0 to 100 0
\endpicture
$$
\caption{\figsize\sf\label{fig-1}A disappearance process.}\end{figure}}

We begin by describing our analog Banks-Zaks model, and
its solution. It is a 2D model of massless fermions coupled to a massive
vector field. We call it the Sommerfield model because it was solved by
Sommerfield\footnote{Georgi's PhD advisor and Schwinger's student.} in
1963~\cite{Sommerfield:1963}.
Next, we describe the high energy
physics that couples the Sommerfield model to our toy standard model, which
is simply a massive scalar carrying a global $U(1)$ charge. In the
infrared, the resulting interaction flows to a coupling of two charged
scalars to an unparticle field with a non-trivial anomalous dimension.
We apply the operator product expansion to
the solution of the Sommerfield model to find the correlation functions of
the low energy unparticle operator.
Finally, we study the simplest unparticle process
shown in figure~\ref{fig-1} in
which two toy standard model scalars ``disappear'' into unparticle stuff.
Because we have the exact solution for the unparticle correlation functions,
we can see precisely how the system makes the
transition from low energy unparticle physics
to the high energy physics of free particles. The answer is simple and
elegant. The ``spectrum'' of the model consists of unparticle stuff and massive bosons. As the incoming energy of the standard model scalars is increased, the unparticle stuff is always there
but more and more massive bosons are emitted and the combination
becomes more and more like the free fermion cross section.

The Sommerfield(-Thirring-Wess)
model~\cite{Sommerfield:1963,Thirring-Wess:1964} is the
Schwinger model~\cite{Schwinger:1962tp} with an additional mass term for
the vector boson:\footnote{Our conventions are $g^{00}=-g^{11}=1,\;
\gamma^0=
\pmatrix{
0&1\cr
1&0\cr
},\;
\gamma^1=
\pmatrix{
0&-1\cr
1&0\cr
},\;
\gamma^5=\gamma^0\gamma^1=
\pmatrix{
1&0\cr
0&-1\cr
}\,$.}
\be
\mathcal{L}=
\bar\psi\,(i\spur\partial - e\,\slash A)\,\psi
-\frac{1}{4}F^{\mu\nu}F_{\mu\nu}
+\frac{m_0^2}{2}A^\mu A_\mu
\label{Sommerfield-model}
\ee
We are interested in this theory since it is exactly solvable and (unlike
the Schwinger model) has fractional anomalous dimensions. In particular, we
are interested in the composite operator,
\begin{equation}
{\cal O}\equiv \psi_2^* \psi_1
\label{o}
\end{equation}
because in the low-energy theory, it scales with an anomalous dimension.

The solution for all fermion Green's functions in the model can be written
down explicitly, in terms of propagators for free fermions, and for
massless and massive scalar fields with mass $m$,
\be
m^2 = m_0^2 + \frac{e^2}{\pi}
\ee
The physical mass $m$ plays the role in this model of the unparticle scale
$\Lambda_\un$ from~\cite{Georgi:2007ek},
setting the scale of the transition between free particle behavior at high
energies and unparticle behavior at low energies.
Explicitly, the scalar propagators are\footnote{$K_0$ is the modified
Bessel function of the
second kind and $x_0$ is an arbitrary constant that will cancel out in the
following. For $y \to \infty$, $K_0(y) \sim \sqrt\frac{\pi}{2y}\,e^{-y} \to
0$. For $y \to 0$, $K_0(y) = -\ln(y/2) - \gamma_E + \cO(y^2)$, where
$\gamma_E = -\Gamma'(1) \simeq 0.577$ is Euler's constant. Note that $\Delta(0)-D(0)=(i/2\pi)\ln\left(e^{\gamma_E}x_0m/2\right)$ is finite.}
\bea
&&\Delta(x) = \int\frac{d^2p}{(2\pi)^2}\frac{e^{-ipx}}{p^2 - m^2 + i\epsilon}
= -\frac{i}{2\pi}K_0\left(m\sqrt{-x^2 + i\epsilon}\right)
\label{massless} \\
&&D(x) = \int\frac{d^2p}{(2\pi)^2}\frac{e^{-ipx}}{p^2 + i\epsilon} = \frac{i}{4\pi}\ln\left(\frac{-x^2+i\epsilon}{x_0^2}\right)
\label{massive}
\eea

The $n$-point functions for the $\cO$ and $\cO^*$ fields
can then be constructed using the operator product expansion. We will
describe all this in detail in a separate
publication~\cite{kats}.\footnote{Here we will also include more complete
references to the unparticle and Sommerfield model literature.}
Here we will simply
write down and use the result for the $2$-point function in position space,
\be
i\Delta_{{\cal O}}(x) \equiv
\bigl\langle0\bigr|\mbox{T}{\cal O}(x)
\,{\cal O}^*(0)\bigl|0\bigr\rangle
= \frac{B(x)}{4\pi^2\left(-x^2 + i\epsilon\right)}
\label{psi-psi-prop}
\ee
where
\bea
B(x)&=& \exp\left[i\frac{4e^2}{m^2}\left[(\Delta(x) - \Delta(0)) - (D(x) - D(0))\right]\right] \nn\\
&=&\exp\left[\frac{2e^2}{\pi m^2}
\left[K_0\left(m\sqrt{-x^2 + i\epsilon}\right) + \ln\left(\xi
m\sqrt{-x^2 + i\epsilon}\right)\right]\right]
\label{b}
\eea
with
\begin{equation}
\xi = e^{\gamma_E}/2
\label{kappa}
\end{equation}

In the short-distance limit ($\left|x^2\right| \ll 1/m^2$), $B(x)\to1$
and one obtains free-fermion behavior. In the large-distance limit
($\left|x^2\right| \gg 1/m^2$), $K_0$ does not contribute so $B(x)$
is just a power of $x^2$, and the 2-point function is proportional to
an unparticle propagator\footnote{Here and below, we incorporate a
dimensional factor of $1/(\xi m)^{2a}$ in the unparticle propagator so
it matches smoothly onto the ${\cal O}$ propagator.}
\be
i\Delta_{{\cal O}}(x) \;\to\; i\Delta_\un(x) =
\frac{1}{4\pi^2(\xi m)^{2a}\left(-x^2 + i\epsilon\right)^{1+a}}
\label{un-prop-position}
\ee
where
\be
a \equiv -\frac{e^2}{\pi m^2} = -\frac{1}{1+\pi m_0^2/e^2}
\label{anom-dim-5}
\ee
Thus at large distance and low energies, the composite operator ${\cal O}$
scales with dimension $1+a$, corresponding to an
anomalous dimension of $a$ for $\psi_2^*\psi_1$. For $0<m_0<\infty$,
$a$ is
fractional, which leads to unparticle behavior. In momentum space
\be
i\Delta_\un(p)
= \frac{iA(a)}{2(\xi m)^{2a}\,\sin(\pi a)}(-p^2 - i\epsilon)^a
= \frac{A(a)}{2\pi(\xi m)^{2a}}\int_0^\infty
dM^2\left(M^2\right)^a\frac{i}{p^2 - M^2 + i\epsilon}
\label{un-prop}
\ee
where the function
\be
A(a) \equiv -\frac{\sin(\pi a)\,\Gamma(-a)}{2^{1+2a}\,\pi\Gamma(1+a)}
\ee
is positive in the range relevant to our model ($-1 < a < 0$). Since
\be
\mbox{Im}\,\Delta_\un(p) = -\frac{A(a)}{2(\xi
m)^{2a}}\theta(p^2)\left(p^2\right)^a
\label{Im-un-prop}
\ee
the unparticle phase space is
\be
\Phi_\un(p) = \frac{A(a)}{(\xi m)^{2a}}\,
\theta(p^0)\,\theta(p^2)\left(p^2\right)^a
\label{un-phase-space}
\ee

To generate a coupling to a toy standard model\label{standard},
we assume that the very high energy theory includes the interaction
\bea
\mathcal{L}_{\rm int} &=& \frac{\mu}{2}\,
\left[\bar\psi(1+\gamma_5)\chi\,\phi^*
+
\bar\psi(1-\gamma_5)\chi\,\phi\right]
+ \mbox{h.c.}\nn\\
&=& \mu\,
\left(\psi_2^*\,\chi_1\,\phi^*
+
\psi_1^*\,\chi_2\,\phi\right)
+ \mbox{h.c.}
\label{highenergy}
\eea
that couples the fermion $\psi$ of the unparticle sector to a neutral
complex scalar $\phi$ with mass $m_\phi \ll m$ that plays the role of a
standard model field. The interaction is mediated by the heavy fermion
$\chi$ with mass $M \gg m,\, \mu^2/m$ and the same coupling to $A^\mu$ as
$\psi$. The theory has a global $U(1)$ symmetry with charge $+1$ for
$\phi^*$, $\psi_1^*$ and $\psi_2$, and $-1$ for $\phi$, $\psi_2^*$ and
$\psi_1$, and $0$ for $\chi$. Integrating out $\chi$ we obtain
\be
\mathcal{L}_{\rm int} =
\frac{h}{2}
\left({\cal O}\,{\phi^*}^2 + {\cal O}^*\phi^2\right)\,,\qquad
h \equiv \frac{2\mu^2}{M}
\label{lowenergy}
\ee
The composite operator $\cO$ defined in (\ref{o})
has charge $-2$ under the global $U(1)$ symmetry.

In a 4D unparticle theory, the interaction corresponding to
(\ref{lowenergy}) would typically be nonrenormalizable, becoming more
important as the energy increases. That does not happen in our 2D toy model. But
we can and
will study the process in figure~\ref{fig-1} in the unparticle limit, and
learn something about the transition region between the ordinary particle
physics behavior at energies large compared to $m$
and the unparticle physics at low energies.

To that end, we consider the physical process
$\phi + \phi \to \mbox{Sommerfield stuff}$ shown in figure~\ref{fig-1}:
Because $\phi^2$ couples to ${\cal O}^*$,
we can obtain
the total cross-section for this process from the discontinuity across the
physical cut in the ${\cal O}$ 2-point function. This is analogous to
the optical theorem for ordinary particle production. For $\phi$ momenta
$P_1$  and $P_2$, this is
\be
\sigma = \frac{\mbox{Im}\,{\cal M}(P_1,P_2\to P_1,P_2)}{s}
\label{optical}
\ee
with $s = P^2$, $P = P_1 + P_2$, and
\be
i{\cal M}(P_1,P_2\to P_1,P_2) = -ih^2 \Delta_{{\cal O}}(P)
\label{amp22}
\ee
In the unparticle limit ($\sqrt{s} \ll m$), using (\ref{Im-un-prop}), or directly the phase space (\ref{un-phase-space}), we find the fractional power behavior expected with unparticle production:
\be
\sigma = \frac{A(a)}{2}\frac{h^2}{(\xi m)^{2a}}\frac{1}{s^{1-a}}
\ee
On the other hand, in the free particle limit appropriate for high energies
$\sqrt{s} \gg m$, we have $B(x) \to
1$ in (\ref{psi-psi-prop}), and then
\be
\sigma = \frac{h^2}{4}\frac{1}{s}
\label{sigma-short-dist}
\ee
which is the cross-section for $\phi + \phi \to \bar\psi_2 +
\psi_1$.

Since have the exact solution, we can study the transition between the two limits by writing (\ref{psi-psi-prop}) for arbitrary $x$ as
\be
i\Delta_{{\cal O}}(x)
= i\Delta_\un(x)\exp\left[-4\pi ia\Delta(x)\right]
= i\Delta_\un(x)\sum_{n=0}^\infty\frac{\left(-4\pi
a\right)^n}{n!}\left[i\Delta(x)\right]^n
\label{psi-psi-prop-exp}
\ee
At distances not large compared to $1/m$, the higher terms in the
sum in (\ref{psi-psi-prop-exp}) become relevant. Notice that in (\ref{psi-psi-prop-exp}), we have expanded in $a$ only the
terms involving the massive boson propagator. This is critical to our
results. It would be a mistake to expand $i\Delta_\un$ in powers of $a$.
This would introduce spurious infrared divergences because the massless
boson propagator is sick in 1+1 dimensions.\footnote{This statement has a
long history in the mathematical physics literature, going back at least to
\cite{Schroer:1963gw}. For an early summary in English, see
\cite{Tarski:1964}. See also  \cite{Coleman:1973ci}. It is also worth
noting that \cite{Schroer:1963gw} introduces the notion of
``infraparticles'' -- an approach to continuous mass representations of the
Poincare group, which of course includes unparticle stuff. See
\cite{Schroer:2008gd} where we learned of this interesting early
reference.} The model describes not massive and massless bosons,
but rather massive bosons and unparticle stuff. In momentum space we obtain
\bea
i\Delta_{{\cal O}}(P)
= \sum_{n=0}^\infty\frac{\left(-4\pi
a\right)^n}{n!}\int\frac{d^2p_\un}{(2\pi)^2}
\,i\Delta_\un(p_\un)\left[\prod_{i=1}^n\frac{d^2p_i}{(2\pi)^2}
i\Delta(p_i)\right](2\pi)^2\delta^2\left(P-p_\un-\sum_{j=1}^n
p_j\right)\nn\\
\eea
This describes a sum of two-point diagrams in which the incoming momentum
$P$ splits between the unparticle propagator and $n$ massive scalar
propagators. Each $\Delta$
is associated with the propagation of a free\footnote{in the absence of the interactions with the $\phi\,$s which are treated perturbatively.} massive
scalar field, so this gives the discontinuity
\be
\begin{array}{c}
\displaystyle
\Phi(P) = \frac{A(a)}{(\xi m)^{2a}}\sum_{n=0}^\infty\frac{\left(-4\pi
a\right)^n}{n!}
\int\frac{d^2p_\un}{(2\pi)^2}\theta(p_\un^0)
\theta(p_\un^2)\left(p_\un^2\right)^a\\ \displaystyle
\times\left[\prod_{i=1}^n\frac{d^2p_i}{(2\pi)^2}\,
\times 2\pi\delta(p_i^2-m^2)\theta(p_i^0)\right]
\times(2\pi)^2\delta^2\left(P - p_\un - \sum_{j=1}^n p_j\right)
\end{array}
\label{full-phase-space}
\ee

For $\sqrt{s}<Nm$, only the first $N$ terms in (\ref{full-phase-space}) (those
involving the production of fewer than $N$ massive bosons) contribute and
(\ref{amp22}) describes the production of unparticle
stuff plus between $0$ and
$N-1$ massive bosons.
For $\sqrt{s}<m$, we have pure unparticle behavior. As we go to higher
energies, the unparticle stuff is always present,
but the emission of more and more massive
bosons builds up the inclusive result for free fermion production. This
happens quickly if $a$ is small, but very gradually for $a$ close to
$-1$.

One can easily obtain explicit results in the case of small $a$, when only
the first few terms in the expansion contribute. The leading correction in
$a$ comes from $n=1$:
\be
\Phi^{(1)} = -a\,\theta(\sqrt{s}-m)\ln\frac{\sqrt{s}}{m} + {\cal O}(a^2)
\ee
which gives the total phase space as
\bea
\Phi = \frac{1}{2} - a\left[\ln\left(\frac{2}{e^{\gamma_E}}\frac{\xi
m}{\sqrt{s}}\right) + \theta(\sqrt{s}-m)\,\ln\frac{\sqrt{s}}{m}\right] + \cO(a^2)
\label{transition}
\eea
For energies $\sqrt{s} > m$, this expression reduces to
\be
\Phi = \frac{1}{2} + \cO(a^2)
\ee
that is the free-fermion result (\ref{sigma-short-dist}). Thus, for $|a|
\ll 1$ there is a discontinuity in $d\Phi/d\sqrt{s}$ at $\sqrt{s} = m$, where a
transition occurs from pure unparticle behavior below energy $m$ to pure
free-fermion behavior above $m$ (see figure~\ref{fig-5}).\footnote{The
linear approximation (\ref{transition}) is not valid for $\sqrt{s} \ll m$ due to
large $\ln \sqrt{s}$, but we have the exact expression (\ref{un-phase-space}).}
To this order, the free-fermion behavior is a sum of the unparticle and the
massive scalar contributions.
{\figsize
\begin{figure}[htb]
$$\epsfxsize=2in \epsfbox[70 250 540 720]{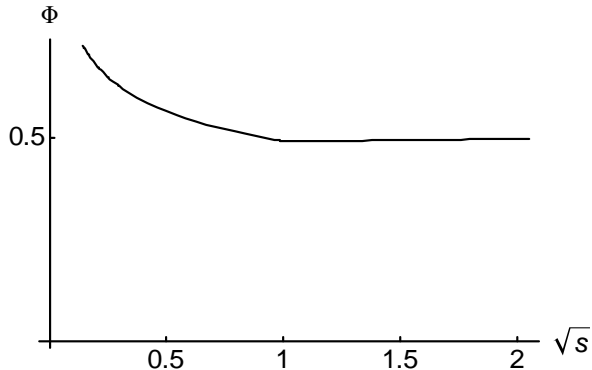}$$
\caption{\figsize\sf\label{fig-5}Phase space $\Phi$ for the disappearance
process in figure~\ref{fig-1} as a function of the energy $\sqrt{s}$ (in units of $m$) for $a = -0.1$.}
\end{figure}}

For larger values of $|a|$, higher powers of $a$ must be included
in (\ref{transition}) to approximate the free-fermion
regime. Since each new massive scalar gives a contribution
with only one additional power of $a$, if $a$ is close to $-1$, the free
fermion behavior is approached very slowly. In fact, the limit $a\to-1$ is singular, and it corresponds to the Schwinger model ($m_0=0$). As is often the case, it is not trivial to obtain a gauge theory as the limit of a theory with a massive vector boson. The unparticle stuff is absent since $A(-1)=0$, and the spectrum includes only a massive boson with $m^2 = e^2/\pi$. The case of the Schwinger model has been studied in~\cite{Casher:1974vf}.

We find this picture of the unparticle scale $\Lambda_\un = m$ in the Sommerfield model very
satisfying. There is a close analog between the way $m$ enters in the process
of figure~\ref{fig-1}, and the way the dimensional transmutation scale
$\Lambda_{\rm QCD}$ enters in inclusive processes in QCD. In QCD, the
physical states are hadrons, typically with masses of the order of
$\Lambda_{\rm QCD}$ unless they are protected by some symmetry (like the
pions). But in the total $e^+e^-$ cross-section into hadrons (to pick the
simplest and most famous example) at high energy $E$, the sum over physical
states  reproduces
the ``parton model'' result with calculable  corrections of order
$1/\ln(E/\Lambda_{\rm QCD})$. We have shown that the process  of
figure~\ref{fig-1} in the Sommerfield model works the same way, with the
physical states being the massive boson and unparticle stuff. Note however,
that in the presence of the standard model couplings, the massive boson is
unstable, decaying into $\phi^*+\phi^*+\mbox{unparticle stuff}$  or
$\phi+\phi+\mbox{anti-stuff}$.
The rate is of order $h^2a^2/m$ so this process is very
slow for small $a$.  We will discuss this further in~\cite{kats}.

\section*{Acknowledgments}

We are grateful to
C. Cordova,
V. Lysov,
P. Petrov,
A. Sajjad, and
D. Simmons-Duffin,
for discussions.
This research is supported in part by
the National Science Foundation under grant PHY-0244821.

\bibliography{up3}

\providecommand{\href}[2]{#2}\begingroup\raggedright\begin{thebibliography}{10}

\bibitem{Georgi:2007ek}
H.~Georgi, ``{Unparticle Physics},'' {\em Phys. Rev. Lett.} {\bf 98} (2007)
  221601,
\href{http://www.arXiv.org/abs/hep-ph/0703260}{{\tt hep-ph/0703260}}.

\bibitem{Georgi:2007si}
H.~Georgi, ``{Another Odd Thing About Unparticle Physics},'' {\em Phys. Lett.}
  {\bf B650} (2007) 275--278,
\href{http://www.arXiv.org/abs/0704.2457}{{\tt 0704.2457}}.

\bibitem{Banks:1981nn}
T.~Banks and A.~Zaks, ``{On the Phase Structure of Vector-Like Gauge Theories
  with Massless Fermions},'' {\em Nucl. Phys.} {\bf B196} (1982)
189--204.

\bibitem{Sommerfield:1963}
C.~M. Sommerfield, ``{On the Definition of Currents and the Action Principle in
  Field Theories of One Spatial Dimension},'' {\em Annals Phys.} {\bf 26}
  (1963) 1--43.

\bibitem{Thirring-Wess:1964}
W.~E. Thirring and J.~E. Wess, ``{Solution of a Field Theoretical Model in One
  Space- One Time Dimension},'' {\em Annals Phys.} {\bf 27} (1964) 331--337.

\bibitem{Schwinger:1962tp}
J.~S. Schwinger, ``{Gauge Invariance and Mass. 2},'' {\em Phys. Rev.} {\bf 128}
  (1962)
2425--2429.

\bibitem{kats}
H.~Georgi and Y.~Kats, ``{Nontrivial Unparticle Interactions in 2D},'' {\em
  manuscript in preparation}.

\bibitem{Schroer:1963gw}
B.~Schroer, ``{Infraparticles in quantum field theory},'' {\em Fortsch. Phys.}
  {\bf 11} (1963)
1--31.

\bibitem{Tarski:1964}
J.~Tarski, ``{Representation of Fields in a Two-Dimensional Model Theory},''
  {\em J. Math. Phys.} {\bf 5} (1964) 1713--1722.

\bibitem{Coleman:1973ci}
S.~R. Coleman, ``{There are no Goldstone bosons in two-dimensions},'' {\em
  Commun. Math. Phys.} {\bf 31} (1973)
259--264.

\bibitem{Schroer:2008gd}
B.~Schroer, ``{A note on Infraparticles and Unparticles},''
\href{http://www.arXiv.org/abs/0804.3563}{{\tt 0804.3563}}.

\bibitem{Casher:1974vf}
A.~Casher, J.~B. Kogut, and L.~Susskind, ``{Vacuum polarization and the absence
  of free quarks},'' {\em Phys. Rev.} {\bf D10} (1974)
732--745.

\end{thebibliography}\endgroup

\end{document}